\documentclass[3p,twocolumn,12pt]{elsarticle}

\usepackage{graphicx}

\usepackage{amssymb}

\usepackage{gensymb}

\journal{and accepted by Space Policy; currently in press.}

\begin{document}

\begin{frontmatter}



\title{Scientific Return of a Lunar Elevator}


\author[1]{T.M. Eubanks}
\ead{tme@asteroidinitiatives.com}
\author[1]{C.F. Radley}
\ead{cfr@asteroidinitiatives.com}

\address[1]{Asteroid Initiatives LLC, P.O. Box 141, Clifton Virginia 20124}

\begin{abstract}
The concept of a space elevator dates back to Tsilokovsky, but they are not commonly considered in near-term plans for space exploration, perhaps because a terrestrial elevator would not be possible without considerable improvements in tether material. A Lunar Space Elevator (LSE), however, can be built with current technology using commercially available tether polymers. This paper considers missions leading to infrastructure capable of shortening the time, lowering the cost and enhancing the capabilities of robotic and human explorers.  These missions use planetary scale tethers, strings many thousands of kilometers long stabilized either by rotation or by gravitational gradients.    These systems promise major reduction in transport costs versus chemical rockets, in a rapid timeframe, for a modest investment.   Science will thus benefit as well as commercial activities. 
\end{abstract}


\end{frontmatter}



\section{Introduction}
\label{sec:Introduction}

The long term exploration and development of space would greatly benefit from the use of planetary-scale tethers, both as dynamic tools and for space elevators \cite{Moravec-1977-a,Swan-et-al-2013-a,Swan-2015-a}.

Free-flying tethers must rotate to stay in tension.   Those that rotate so as to cancel the relative motion between the tip and a planetary or satellite surface are called rotovators \cite{Forward-1991-a}; such tethers may be used to set up transportation systems moving material to and from planetary surfaces at low relative velocities and without the expenditure of fuel \cite{Moravec-1977-a,Hoyt-2000-a}.  A space elevator is a tether deployed as a static or orbiting structure stretching from a celestial body out into space \cite{Swan-et-al-2013-a}. In order for a space elevator to remain static (stationary with respect to the surface of the body it is attached to) its center of mass must be in a stationary orbit, with the force of gravity on the tether being balanced by either the centrifugal force of rotation (for a terrestrial elevator) \cite{Aravind-2007-a} or tidal forces (for a lunar elevator) \cite{Pearson-et-al-2005} on the mass of the tether plus any counterweight above the center of mass.

The proposed Deep Space Tether Pathfinder (DSTP)  mission 
is intended to both test the technology of the prototype LSE and provide a substantial scientific return by doing touch-and-go sampling of a selected area on the lunar surface. The rotation of the DSTP would be used to match the relative velocity between its lower tip and the Moon during a flyby, allowing for the collection of surface samples from a suitable scientific target, in the default mission from the floor of Shackleton Crater in the lunar South polar region. The collected material would then be returned to Earth by the release of a return capsule roughly one half rotation period later, when elevator tip velocity is appropriate for a direct return trajectory. After sample release, the DSTP would continue into deep space, allowing for long term observations of the performance and micrometeorite resistance of the tether in the space environment and the first test of kilometric radio interferometry in deep space. 

The proposed LSE Infrastructure (LSEI), the first true space elevator on any celestial body, is planned as  a follow-on to the DSTP. The LSEI would be a very long tether extending from the lunar Surface, through the Earth-Moon Lagrange L1 point (EML-1) 56,000 km above the Moon,  and on into cis-lunar space. The LSEI prototype, scaled to be deployable with one launch of a heavy lift vehicle, would be able to lift roughly 5 tons of lunar samples per year, and deploy a similar quantity of equipment onto the lunar surface. The LSEI would enhance a crewed Deep Space Habitat (DSH) at EML-1, for a small fraction of the total DSH cost by, for example, supporting tele-robotic exploration on the surface. Similar scientific work could be accomplished by a farside LSE, which could also provide real time communications to the farside, opening an entire lunar hemisphere to exploration.

Of the possible near-term space elevator deployments (Earth, Moon, Mars), a lunar nearside elevator is undoubtedly the most technically feasible. 
Modern high strength polymers such as Ultra-high-molecular-weight polyethylene (UHMWPE) (brand name Dyneema\textregistered) \cite{Stein-1998-a} and poly-phenylenebenzobisoxazole (PBO) (brand name Zylon\textregistered) \cite{Wolfe-1988-a} are inexpensive and available in large quantities, ample for LSE tethers.  Even a prototype LSE, deployed by a single launch of an existing launch vehicle, would serve as the linchpin in a lunar delivery service, the LSEI, capable of transporting up to 5 tonnes of material to and from the lunar surface per year and supporting a wide variety of scientific research, including on and near the lunar surface, at the L1 Lagrange point, and deep into cislunar space at the counterweight.      

The $\Delta$V required for a rocket to ascend
from lunar surface to EML-1 is 2.7 km s$^{-1}$. 
Goff \cite{Goff-2013-a} showed that the typical 
payload mass fraction for such a rocket is 34\%,
$\sim$1/3. A rocket which puts the 49 tonnes
LSE at EML-1 would otherwise be capable of
depositing 16 tonnes on to the lunar
surface. So for LSE payload of 0.1 tonnes,
this is equivalent to 16/0.1 = 160 payload
landing cycles, which is the number of cycles
to recoup the LSE launch cost. For sample
return, another factor of three applies, so
$\sim$53 sample return cycles would
recoup the launch cost.

While the initial LSEI would not be able to deliver human passengers to and from the lunar surface, a functioning LSEI prototype would enhance the capabilities of humans in a Deep Space Habitat (DSH) in a Lissajous orbit around EML-1, as envisioned in the  2011 Global  Exploration Roadmap \cite{GER-2013,Crawford-2014-a}. The LSEI would: enable astronauts to deliver rovers and instruments to the lunar surface, teleoperate that equipment from only 56,000 km altitude, lift selected surface samples to EML-1, evaluate those samples, and use that evaluation to direct the acquisition of further samples.

\section{The Scientific Goals of the Deep Space Tether Pathfinder}
\label{sec:DSTP-Goals}

The DSTP would be spacecraft with a 5000 kilometer long tether, with a tether mass of 2228 kg and a total system mass of 3043 kg,
rotating every 2.44 hours with a sampling probe on the far tip \cite{Eubanks-2012-a}.
 The DSTP would flyby the Moon as a rotovator \cite{Forward-1991-a} to collect lunar samples in a touch-and-go manner, followed by a cruise in deep space as an engineering test of the tether technology needed for the first LSE \cite{Eubanks-2012-a}. The DSTP would be the first tether actually deployed as a rotovator, rotating to match the velocity of its sampling tip with the lunar surface, which would enable sample acquisition from a scientifically interesting region, such as the permanently shadowed regions at the lunar poles. Approximately 2 hours after sample collection the DSTP would use its rotational velocity to sling-shot the sample back to Earth for a ballistic reentry with a minimal expenditure of fuel. The DSTP would then continue on into deep space for a long-duration exposure test of the radiation and micrometeorite resistance of the tether's design, and also a test of kilometric radio interferometry in deep space \cite{Eubanks-2012-a}.

The primary scientific justification of the DSTP mission would be lunar sample return; its lunar science objectives address every one of goals in the ``Lunar Polar Volatiles and Associated Processes'' white paper submitted to the 2011 Decadal Survey \cite{NRC-Decadal-2011}.  Current DSTP mission planning has focused on sampling volatiles on the shadowed floor of Shackleton Crater at the lunar South Pole, which is a cold-trap and should collect substantial amounts of surface volatiles from collisions and out-gassing on other areas of the Moon \cite{Paige-et-al-2010-a}. 
Near-surface imagery returned during the sample collection process will help to assess the nature and distribution of volatiles, even if sample return is not successful. (Portions of the Shackleton Crater rim are in sunlight at any time of month \cite{Noda-et-al-2008-a}, providing illumination of the crater bottom that is typically several times full-Moon illumination on Earth.) The search for lunar volatiles ranks high in the decadal surveys of planetary science \cite{NRC-Decadal-2011}, and the Permanently Shadowed Regions (PSRs) on the Moon are arguably the easiest such locations to access in the solar system. The PSRs contain an important scientific record of the history of volatiles in the inner Solar System, and a potential resource for future economic development \cite{Crawford-2015-a,Spudis-2016-a}. These regions have been the target of intense scientific interest in the last decade, and were the target of the LCROSS impactor \cite{Heldmann-et-al-2012-a}, but surface sampling by landers or rovers is complicated by the lack of solar power and direct communications with Earth in a PSR.

\begin{figure}
\centering
\includegraphics[scale=0.55]{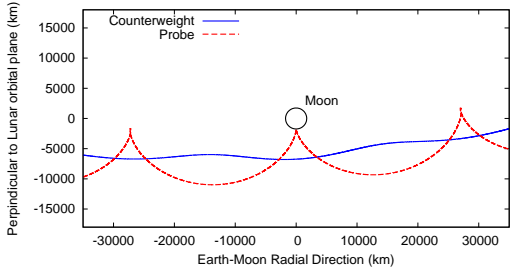}
\caption{Trajectories of the two tips of the DSTP during the entire lunar sample return period, as seen from a selenocentric reference frame \cite{Eubanks-2012-a}. The main spacecraft, assumed to include the upper stage as a counterweight, is considerably more massive than the probe and is thus closer to the tether center-of-mass, which executes a smooth ballistic motion. This Figure represents 6 hours of total motion.}
\label{fig:DSTP-path-1}
\end{figure} 

\begin{figure}
\centering
\includegraphics[scale=0.55]{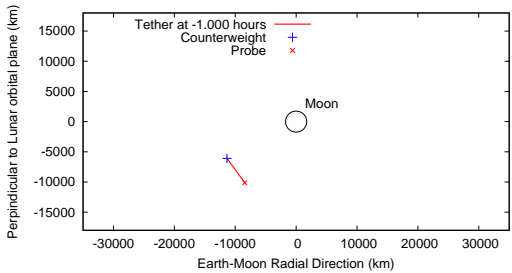}
\caption{The DSTP one hour before the touch-and-go sampling, from the simulation shown in Figure \ref{fig:DSTP-path-1}. }
\label{fig:DSTP-minus-1-hour}
\end{figure}

\begin{figure}
\centering
\includegraphics[scale=0.55]{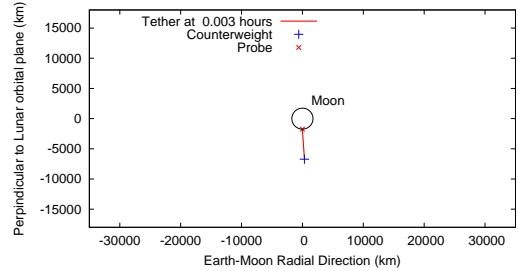}
\caption{The DSTP 10 seconds after the time of the touch-and-go sampling, from the simulation shown in Figure \ref{fig:DSTP-path-1}.}
\label{fig:DSTP-at-contact}
\end{figure}

\begin{figure}
\centering
\includegraphics[scale=0.55]{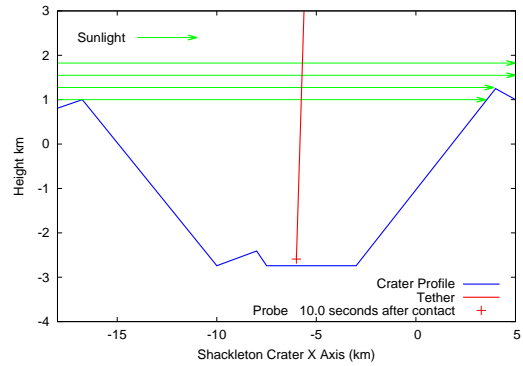}
\caption{A cross sectional view of Shackleton Crater from Lidar data \cite{Smith-et-al-2012-a}, with the DSTP tether and probe 10 seconds after closest approach, when it is $\sim$100 m above the crater floor \cite{Eubanks-2012-a}. This image is from the same simulation sampled at the same time as for Figure \ref{fig:DSTP-at-contact}. The green horizontal arrows indicate the maximum  illumination of the crater by the Sun; the crater interior below these lines is permanently shadowed (the sampling probe is in shadow for $\sim$2 minutes).}
\label{fig:Shackleton_X_center_1}
\end{figure}

Figure \ref{fig:DSTP-path-1} shows the general DSTP trajectory near the Moon in a 2-body gravitational simulation, while Figures \ref{fig:DSTP-minus-1-hour} and \ref{fig:DSTP-at-contact} show the DSTP tether positions one hour before and just after the time of sampling, respectively. Figure \ref{fig:Shackleton_X_center_1}, an enlargement of Figure \ref{fig:DSTP-at-contact} (inverted so that the crater floor is at the bottom), shows that the tether descends almost vertically at the lunar surface; to a surface observer the motion of the probe up and down inside the crater would appear to be almost completely vertical, enabling sampling from topographically rough regions. In addition, there is a clear line-of-sight back to the main spacecraft at the other end of the tether, allowing for direct relay communication with Earth at the time of sampling. 

Shackleton Crater sits on the boundary of the older and much larger South Pole-Aitken Basin, an $\sim$2500-km diameter impact basin which brought up material from deep inside the lunar interior. Given the surface albedo results from Selene \cite{Noda-et-al-2008-a}, it is highly likely that a sample collection from the floor of Shackleton Crater would include rock or regolith samples from the South Pole-Aitken Basin \cite{Spudis-et-al-2008-a}. 

A proposed South PoleÐAitken Basin Sample-Return (SPA-SR) mission was highly ranked by the National Research Council Planetary Science Decadal Survey \cite{NRC-Decadal-2011} and was suggested for a ÒNew FrontiersÓ class mission, with a cost cap of \$1.0 billion. The DSTP would provide a first look at both South Pole-Aitken Basin material, and at the volatiles in a PSR, within a NASA Discovery mission cost cap.

Another scientific goal of the DSTP is to deploy the first  radio interferometer for the kilometric spectral region between 10 kHz and 1 MHz, which is largely unexplored for radio astronomy as these wavelengths do not penetrate the EarthÕs ionosphere. The proposed radio interferometer, the Dark Ages Pathfinder (DAP), would consist of two 10-km dipoles attached to either tip of the DSTP, providing an interferometric baseline of $\sim$5000 km and allowing for rotational synthesis as the tether rotates. At 1 MHz this baseline would allow for an angular resolution of approximately 1\degree, allowing detection of candidate point sources and limited mapping of extended sources. The DAP would complement the proposed Dark Ages Radio Explorer (DARE) \cite{Burns-et-al-2012-a}, which is to operate over a higher frequency radio bandpass of 40-120 MHz. The DAPÕs ability to distinguish solar system, galactic and cosmological sources from terrestrial interference will improve as the distance from the Earth increases. Observations with Radio Astronomy Explorer B \cite{Alexander-et-al-1975-a} indicate that at the lunar distance the Earth interference is typically about 2 orders of magnitude above the celestial background. The terrestrial interference should thus decline to a manageable level when the DSTP is $\gtrsim$3 million km from Earth, or $\sim$2 weeks into the extended deep space mission.

\section{The Prototype Lunar Space Elevator Infrastructure}
\label{sec:LSEI}

\begin{figure}
\centering
\includegraphics[scale=0.38]{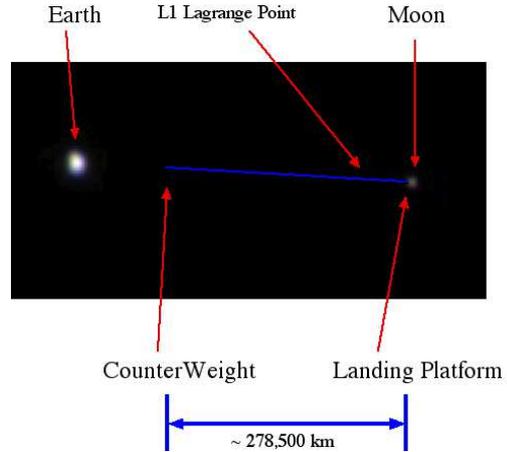}
\caption{The components of the LSEI LSE, to scale, superimposed on a image of the Earth-Moon system from the \textit{Juno} spacecraft.}
\label{fig:LSE-to-scale}
\end{figure}

The LSEI currently is planned to be executed in a single Discovery class mission, starting with the delivery of 58,500 kg of Zylon HM fiber plus associated equipment to the EML-1 Lagrange site.
Figure \ref{fig:LSE-to-scale} shows to scale the major components of LSEI, the string, the Landing Platform (LP), the supply depot at EML-1, and the CounterWeight (CW), while Table \ref{table:2-LSEI} provides basic information about the default LSEI for both lunar hemispheres.

\begin{figure}
\centering
\includegraphics[scale=0.28]{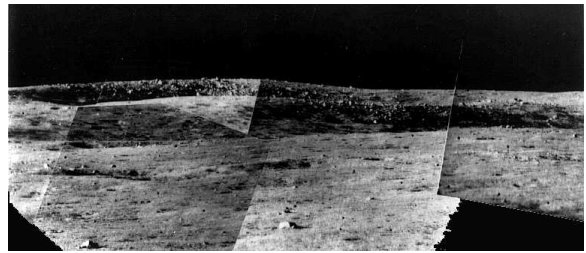}
\caption{Sinus Medii from \textit{Surveyor 6}, taken about 44 km from the proposed landing site (Figure 7-41 from \cite{NASA-Surveyor-1969-a}).}
\label{fig:Sinus-Medii}
\end{figure}

The LP attached to the tether descends to the lunar surface in the initial prototype deployment. After landing, we refer to it as the Landing Station (LS); the planned LS location is Sinus Medii, near 0\degree~Latitude and Longitude on the lunar nearside. Figure \ref{fig:Sinus-Medii} shows the topography of Sinus Medii from the \textit{Surveyor 6} lander \cite{NASA-Surveyor-1969-a}.
There are 3 natural locations for long-term scientific observations from the LSEI, the LS on the surface, the deployment platform at EML-1, and the Counterweight (CW) at the far end of the elevator. All 3 locations should be instrumented, both for the scientific return and to monitor the elevator's performance.

\begin{table*}[t]
	\begin{center}
	\begin{tabular}{l r r}
		\hline
		Parameter &  \multicolumn{2}{c}{Elevator Location} \\
		          & Nearside & Farside \\
		\hline
        		 Tether Material & Zylon PBO & Zylon PBO   \\
		Length  & 278544 km & 297308 km \\
		Mass    & 48700 kg  & 48700 kg \\
		Surface Lift Capacity & 128 kg & 110 kg \\
		Total Taper (max / min area) & 2.49 & 2.49 \\
		Maximum Force & 517 N & 446 N \\
		Landing Site & 0\degree\ E 0\degree\ N & 180\degree\ E 0\degree\ N \\
		\hline
	\end{tabular}
	\end{center}
\caption{Prototype Lunar Elevators \cite{Eubanks-2013-e}.}
\label{table:2-LSEI}
\end{table*}

The primary initial science goal of the LSEI prototype mission is the return of the lunar samples to Earth. LSEI will take a core sample upon landing and will deliver one or more microrovers to the lunar surface to assist in collecting surface samples. LSEI will return up to 100 kg of samples in the first Lift from the lunar surface, using a reusable solar-powered lifter. Sample returns can be done without fuel using a nearside LSE, as material (in a suitable return capsule) could be simply released at the right moment for a direct reentry trajectory to a desired landing location; anything separated from the LSE an altitude $\gtrsim$ 220,670 km above lunar surface will re-enter the Earth's 
atmosphere in $\sim$1.4 days at a velocity of $\sim$10.9 km s$^{-1}$ without any expenditure of fuel. This same technique can be used to return high value ore samples or mining products from a lunar mining enterprise.   

There are some important sites with materials of economic interest near to the EML-1 LS.  The nearby crater Lalande is known to have some of the highest concentrations of surface KREEP deposits on the Moon \cite{Lawrence-et-al-2003-a}, as well as 19 impact melt pits, possible sites for volatiles \cite{Wagner-Robinson-2014-a}.   Also, nearby mare areas appear to have elevated concentrations of Helium-3 \cite{Fa-Jin-2007}.   Regardless of landing site, lowland regolith contains concentrations of lunar volatiles which can be used for propulsion and life support.  The LS will also become an important transit point for long distance lunar rovers which will recover samples over large distances much more cheaply than using rocket landers.

LSEI plans to use Single Cube Retroreflectors (SCR) as Laser ranging targets \cite{Turyshev-et-al-2013-a} for navigation during deployment of the LSE and the Landing Platform. The SCR would become a permanent addition to the Lunar Laser Ranging (LLR) retroreflector network. The LSEI would thus augment LLR studies of basic physics, lunar dynamics, and Earth-Moon celestial mechanics \cite{Williams-et-al-2006-a}. 

Poorly understood electrostatic levitation and transport of dust happens in regions near the lunar terminator \cite{Grun-et-al-2011-a}, and may be important in the covering of PSR volatiles over geologic time. If electrostatically-levitated dust is present LSEI will sample it \textit{in situ} with passive Aerogel collectors or electrets (permanently charged materials), deployed at altitude during periods with a elevator lift is not scheduled.

The CW can observe the Earth from over 100,000 km away, outside of the existing  satellite constellations. It will be able to observe the magnetic and charged particle environment in the EarthÕs magnetopause as the CW goes in and out of the EarthÕs magnetosphere twice per lunar month.

In addition, EML-1 is a logical location for the observation of the nearside of the Moon. One fairly small optical telescope (20 cm) could continuously observe the entire nearside, searching for meteorite impacts and transient lunar phenomena, and also be able to detect and characterize the orbits of close lunar orbiters.

Various factors could limit the useful life of an LSE, with micrometeoroid impacts being an especially serious threat to tether longevity (in cislunar space there is no significant flux of man-made orbital debris).
 The LSEI would be a very thin tether, with a radius of $\sim$0.2 mm if it were just a single strand. Such a strand would be broken if impacted by a meteorite with a mass as small as 10$^{-5}$ gm. A variety of methods have been used to determine the micrometeorite flux in near-Earth and near-lunar space \cite{Grun-et-al-2011-a,Drolshagen-et-al-2008-a}; the cumulative flux of meteorites of this mass and larger is $\sim$10$^{-8}$ m$^{-2}$ s$^{-1}$. The LSEI is sufficiently long that its surface area would be $\sim$10$^{5}$ m$^{2}$; if the LSEI were made from a single strand it would have a micrometeorite impact lifetime measured in hours. The LSEI will have to deploy a multiline fail-safe system such as the ``Hoytether'' \cite{Forward-Hoyt-1995-a} to achieve a design life-time of 5 years; testing of the chosen micrometeorite protection system in deep space would be one of the primary engineering goals of the DSTP mission.

\section{A Farside Lunar Space Elevator}
\label{sec:Far-Side-LSE}

An elevator on the lunar farside (with a landing point at or near longitude 180\degree, latitude 0\degree) could fulfill many of the scientific and logistical goals of a nearside LSEI, but would also provide unique advantages of its own \cite{Eubanks-2013-e,Eubanks-et-al-2015-a}.

\subsection{Sample Return from the Lunar Farside}
\label{subsec:Sample-Return-from-the-Lunar-Farside}

\begin{figure}
\centering
\includegraphics[scale=0.45]{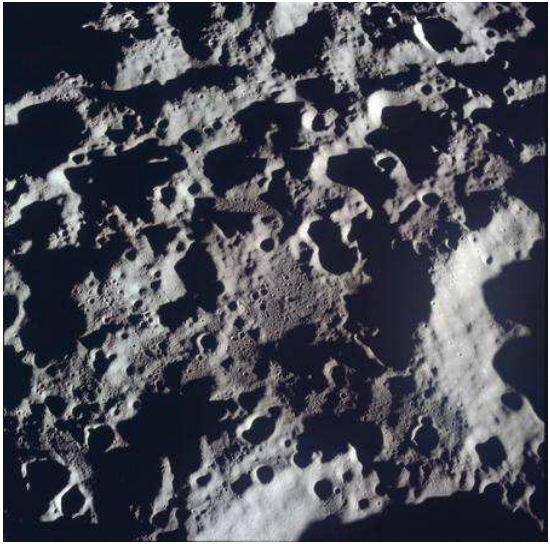}
\caption{Apollo Image AS11-44-6607, taken by astronaut Michael Collins during the \textit{Apollo 11} mission, July, 1969, from an altitude of $\sim$110 km. This image shows the very rough highland terrain at the farside elevator LS (the default LS is towards the upper left of this image). This part of the lunar terrain has never been explored or sampled by any surface mission.}
\label{fig:AS11-44-6607}
\end{figure}

To date, all lunar sample returns have been from from 9 sites on the lunar nearside \cite{Jaumann-et-al-2013-a}. The LSE in Table \ref{table:2-LSEI} assume ÒnaturalÓ elevator landing sites (i.e., directly beneath the Lagrange Point), as these seem most appropriate for a initial elevator deployment. An EML-2 LSE would thus provide an immediate sample return from a previously unsampled region (and, indeed, from a previously unsampled hemisphere).  The EML-2 landing site (Figure \ref{fig:AS11-44-6607}) is near Lipskiy Crater and just North of Daedalus Crater in very rugged and heavily cratered terrain in the lunar highlands.

\subsection{Farside Radio Astronomy}
\label{subsec:Farside-Radio-Astronomy}

The farside of the Moon is totally shielded from terrestrial radio transmissions, and is arguably the best place in the solar system for a radio astronomy base. The Earth is a major source of radio noise and interference, both natural and artificial, and its ionosphere blocks ground based observations at frequencies $\lesssim$ 10 MHz. The farside LSE LS at
180\degree~E longitude point would be an ideal location for such a farside radio astronomy base, far from terrestrial interference and with a view of the entire sky.    Many have proposed radiotelescopes there, e.g. Jester and Falcke \cite{Jester-Falcke-2009-a}; an EML-2 LSEI would considerably reduce the cost of building and supplying a lunar farside radio telescope system, enabling both the installation of antennas on the surface at the LS, as well as  vertically using the lower portion of the elevator as a antenna tower \cite{Eubanks-et-al-2015-a}. Decametric and kilometric radio astronomy could be conducted during the lunar night, when radio interference from the Sun is also blocked and when solar powered climbers would not be using the near surface part of the LSE. 

The farside is recommended as a radio quiet zone by the International Telecommunications Union under ITU-R RA.479. Maccone \cite{Maccone-2008-a} has proposed a more extensive Protected Antipode Circle [PAC], a more extensive protection zone than that proposed by the ITU. The PAC centered around the antipode on the farside spanning an angle of 30¡ in longitude and latitude in all radial directions from the antipode. The PAC is the most shielded area of the farside, with an expected attenuation of man-made RFI of 100 dB or higher.  Neither the PAC nor ITU farside rules  have  been adopted by any law making body; in any case a farside LSE would have to avoid interference with farside radio observatories, whether at the LS or installed elsewhere by other missions.

\subsection{Other Farside Science}
\label{subsec:Other-Farside-Science}

An EML-2 LSE would enable a variety of other farside science, including the monitoring of particles and fields in near interplanetary space at EML-2 and at the far end of the elevator, which would enable deep monitoring along the Earth magnetotail at Full Moon. From a EML-2 station almost the entire farside could be monitored for meteor impacts, complementing the terrestrial monitoring of the nearside for impacts \cite{Suggs-et-al-2014-a}. The monitoring of the time and location of farside impacts will be especially important if a nearside lunar seismological network is re-established, as impacts on the farside will provide seismic waves traversing the lunar core  to nearside seismometers. A EML-2 LSE would extend the lunar seismological network to the farside itself, providing a truly global lunar monitoring network. Even one seismometer at the farside LS would enable the seismic study of the entire lunar interior \cite{Lognonne-et-al-2012-a} in combination with the locations and times of nearside impacts provided  the ongoing terrestrial monitoring of meteorite impacts visible lunar surface \cite{Suggs-et-al-2014-a}.

\subsection{Farside Communications Relay}
\label{subsec:Farside-Communications}

Communications has always been a severe complication for the engineering of missions to the lunar farside, as there is no direct line-of-site between the Earth and any location deep in the farside (librations bring occasional line-of-sight to locations at the farside-nearside boundary). A EML-2 LSE would provide a communications mast visible from almost any location on the lunar farside, and could thus serve as a relay for communications with the Earth.  A farside LSE communications relay would literally open an entire lunar hemisphere for further exploration.  

\section{Facilitation of Mars Exploration}
\label{sec:Mars-exploration}

Ishimatsu \textit{et al.} \cite{Ishimatsu-et-al-2015-a} have shown that the launch weight of missions to Mars can be reduced by 68\% by using oxygen derived from the Moon as propellant.   An LSE would allow  relatively inexpensive transport of lunar regolith to a cislunar oxygen production facility, further reducing transportation costs. Volatiles derived from the lunar regolith could fulfill most of the life support consumables required by a crewed Mars mission. Resources from the Moon can facilitate missions to anywhere in the solar system \cite{Crawford-2015-a}, and lunar elevators could greatly reduce the cost of transporting such materials into space. 

In addition, the velocity of the counterweight of a farside LSE would provide significant $\Delta$V for injection into a trans-Mars orbit. Material lifted past EML-2 could be sent to Mars with a minimal investment in fuel, helping to support long-term colonies on that planet.

\section{Conclusions}
\label{sec:Conclusions}

The DSTP and the LSEI are crucial first steps in the development of space elevators, and in future tether missions, but can and must be justified on the basis of returned science, in addition to their engineering return. Building the DSTP is feasible, has an exciting scientific return and would be natural first step in developing an elevator-based lunar infrastructure program, leading to the construction of the LSEI and in due course the development of a true transportation infrastructure for the inner solar system.

\bibliographystyle{elsarticle-num}
\bibliography{./eubanks}

\begin{thebibliography}{10}
\expandafter\ifx\csname url\endcsname\relax
  \def\url#1{\texttt{#1}}\fi
\expandafter\ifx\csname urlprefix\endcsname\relax\def\urlprefix{URL }\fi
\expandafter\ifx\csname href\endcsname\relax
  \def\href#1#2{#2} \def\path#1{#1}\fi

\bibitem{Moravec-1977-a}
H.~{Moravec}, {A non-synchronous orbital skyhook}, Journal of the Astronautical
  Sciences 25 (1977) 307--322.

\bibitem{Swan-et-al-2013-a}
P.~A. {Swan}, D.~I. {Raitt}, C.~W. {Swan}, R.~E. {Penny}, J.~M. {Knapman},
  {Space Elevators: An Assessment of the Technological Feasibility and the Way
  Forward}, International Academy of Astronautics, 2013.

\bibitem{Swan-2015-a}
P.~A. {Swan}, {Opening up Earth-Moon Enterprise with a Space Elevator}, New
  Space 3 (2015) 213--217.
\newblock \href {http://dx.doi.org/10.1089/space.2015.0025}
  {\path{doi:10.1089/space.2015.0025}}.

\bibitem{Forward-1991-a}
R.~L. {Forward}, {Tether Transport from LEO to the Lunar Surface}, in:
  AIAA/ASMA/SAE/ASEE 27th Joint Propulsion Confererence, 1991, pp.
  AIAA--91--2322.

\bibitem{Hoyt-2000-a}
R.~P. {Hoyt}, {Cislunar Transport System}, in: Proc. 2nd Lunar Development
  Conference, 2000, pp. AIAA--99--2690.

\bibitem{Aravind-2007-a}
P.~K. {Aravind}, {The physics of the space elevator}, American Journal of
  Physics 75 (2007) 125--130.
\newblock \href {http://dx.doi.org/10.1119/1.2404957}
  {\path{doi:10.1119/1.2404957}}.

\bibitem{Pearson-et-al-2005}
J.~Pearson, E.~Levin, J.~Oldson, H.~Wykes, {Lunar Space Elevators for Cislunar
  Space Development}, {Phase I Final Technical Report}, {Star Technology and
  Research, Inc.} (2005).

\bibitem{Stein-1998-a}
H.~L. {Stein}, {Ultrahigh molecular weight polyethylenes (uhmwpe)}, in:
  {Engineered Materials Handbook: Engineering Plastics}, Vol.~2, {ASM
  International}, 1998, pp. 167--171.

\bibitem{Wolfe-1988-a}
J.~F. {Wolfe}, {Polybenzothiazoles and Oxazoles}, in: {Encyclopedia of Polymer
  Science and Engineering}, Vol.~11, {John Wiley and Sons}, 1988, p. 601.

\bibitem{Goff-2013-a}
J.~Goff, The slings and arrows of outrageous lunar transportation schemes: Part
  1–gear ratios, in, the Selenian Boondocks.
  http://selenianboondocks.com/2013/12/the-slings-and-arrows-of-outrageous-lunar-transportation-schemes-part-1-gear-ratios/
  (2013).

\bibitem{GER-2013}
{International Space Exploration Coordination Group}, {The Global Exploration
  Roadmap}, Tech. Rep. NP-2013-06-945-HQ, NASA,
  https://www.globalspaceexploration.org/ (2013).

\bibitem{Crawford-2014-a}
I.~A. {Crawford}, {Introduction to the Special Issue on the Global Exploration
  Roadmap}, Space Policy 30 (2014) 141--142.

\bibitem{Eubanks-2012-a}
T.~M. {Eubanks}, {Sample Return from Shackleton Crater with the Deep Space
  Tether Pathfinder (DSTP)}, in: Lunar and Planetary Science Conference,
  Vol.~43 of Lunar and Planetary Inst.~Technical Report, 2012, p. 2870.

\bibitem{NRC-Decadal-2011}
{National Research Council}, Vision and Voyages for Planetary Science in the
  Decade 2013-2022, National Academies Press, Washington, D.C., 2011.

\bibitem{Paige-et-al-2010-a}
D.~A. {Paige}, M.~A. {Siegler}, J.~A. {Zhang}, P.~O. {Hayne}, E.~J. {Foote},
  K.~A. {Bennett}, A.~R. {Vasavada}, B.~T. {Greenhagen}, J.~T. {Schofield},
  D.~J. {McCleese}, M.~C. {Foote}, E.~{DeJong}, B.~G. {Bills}, W.~{Hartford},
  B.~C. {Murray}, C.~C. {Allen}, K.~{Snook}, L.~A. {Soderblom}, S.~{Calcutt},
  F.~W. {Taylor}, N.~E. {Bowles}, J.~L. {Bandfield}, R.~{Elphic}, R.~{Ghent},
  T.~D. {Glotch}, M.~B. {Wyatt}, P.~G. {Lucey}, {Diviner Lunar Radiometer
  Observations of Cold Traps in the Moon's South Polar Region}, Science 330
  (2010) 479.
\newblock \href {http://dx.doi.org/10.1126/science.1187726}
  {\path{doi:10.1126/science.1187726}}.

\bibitem{Noda-et-al-2008-a}
H.~Noda, H.~Araki, S.~Goossens, Y.~Ishihara, K.~Matsumoto, S.~Tazawa,
  N.~Kawano, S.~Sasaki, Illumination conditions at the lunar polar regions by
  kaguya(selene) laser altimeter, Geophysical Research Letters 35~(24) (2008)
  L24203, l24203.
\newblock \href {http://dx.doi.org/10.1029/2008GL035692}
  {\path{doi:10.1029/2008GL035692}}.

\bibitem{Crawford-2015-a}
I.~A. {Crawford}, {Lunar Resources: A Review}, Progress in Physical Geography
  39 (2015) 137--167.
\newblock \href {http://arxiv.org/abs/1410.6865} {\path{arXiv:1410.6865}},
  \href {http://dx.doi.org/10.1177/0309133314567585}
  {\path{doi:10.1177/0309133314567585}}.

\bibitem{Spudis-2016-a}
P.~Spudis, {The Value of the Moon: How to Explore, Live and Prosper in Space
  Using the Moon's Resources}, Smithsonian Books, Princeton University Press,
  2016.

\bibitem{Heldmann-et-al-2012-a}
J.~L. {Heldmann}, A.~{Colaprete}, D.~H. {Wooden}, R.~F. {Ackermann}, D.~D.
  {Acton}, P.~R. {Backus}, V.~{Bailey}, J.~G. {Ball}, W.~C. {Barott}, S.~K.
  {Blair}, M.~W. {Buie}, S.~{Callahan}, N.~J. {Chanover}, Y.-J. {Choi},
  A.~{Conrad}, D.~M. {Coulson}, K.~B. {Crawford}, R.~{DeHart}, I.~{de Pater},
  M.~{Disanti}, J.~R. {Forster}, R.~{Furusho}, T.~{Fuse}, T.~{Geballe}, J.~D.
  {Gibson}, D.~{Goldstein}, S.~A. {Gregory}, D.~J. {Gutierrez}, R.~T.
  {Hamilton}, T.~{Hamura}, D.~E. {Harker}, G.~R. {Harp}, J.~{Haruyama},
  M.~{Hastie}, Y.~{Hayano}, P.~{Hinz}, P.~K. {Hong}, S.~P. {James},
  T.~{Kadono}, H.~{Kawakita}, M.~S. {Kelley}, D.~L. {Kim}, K.~{Kurosawa}, D.-H.
  {Lee}, M.~{Long}, P.~G. {Lucey}, K.~{Marach}, A.~C. {Matulonis}, R.~M.
  {McDermid}, R.~{McMillan}, C.~{Miller}, H.-K. {Moon}, R.~{Nakamura},
  H.~{Noda}, N.~{Okamura}, L.~{Ong}, D.~{Porter}, J.~J. {Puschell}, J.~T.
  {Rayner}, J.~J. {Rembold}, K.~C. {Roth}, R.~J. {Rudy}, R.~W. {Russell}, E.~V.
  {Ryan}, W.~H. {Ryan}, T.~{Sekiguchi}, Y.~{Sekine}, M.~A. {Skinner},
  M.~{S{\^o}ma}, A.~W. {Stephens}, A.~{Storrs}, R.~M. {Suggs}, S.~{Sugita},
  E.-C. {Sung}, N.~{Takatoh}, J.~C. {Tarter}, S.~M. {Taylor}, H.~{Terada},
  C.~J. {Trujillo}, V.~{Vaitheeswaran}, F.~{Vilas}, B.~D. {Walls}, J.-i.
  {Watanabe}, W.~J. {Welch}, C.~E. {Woodward}, H.-S. {Yim}, E.~F. {Young},
  {LCROSS (Lunar Crater Observation and Sensing Satellite) Observation
  Campaign: Strategies, Implementation, and Lessons Learned}, Space Science
  Reviews 167 (2012) 93--140.
\newblock \href {http://dx.doi.org/10.1007/s11214-011-9759-y}
  {\path{doi:10.1007/s11214-011-9759-y}}.

\bibitem{Smith-et-al-2012-a}
D.~E. {Smith}, M.~T. {Zuber}, J.~W. {Head}, G.~A. {Neumann}, E.~{Mazarico},
  M.~H. {Torrence}, O.~{Aharonson}, A.~R. {Tye}, C.~I. {Fassett}, M.~A.
  {Rosenburg}, H.~J. {Melosh}, {Brightening and Volatile Distribution within
  Shackleton Crater Observed by the LRO Laser Altimeter}, in: Lunar and
  Planetary Science Conference, Vol.~43 of Lunar and Planetary Inst.~Technical
  Report, 2012, p. 1663.

\bibitem{Spudis-et-al-2008-a}
P.~D. {Spudis}, B.~{Bussey}, J.~{Plescia}, J.-L. {Josset}, S.~{Beauvivre},
  {Geology of Shackleton Crater and the south pole of the Moon}, Geophys. Res.
  Lett. 35 (2008) L14201.
\newblock \href {http://dx.doi.org/10.1029/2008GL034468}
  {\path{doi:10.1029/2008GL034468}}.

\bibitem{Burns-et-al-2012-a}
J.~O. {Burns}, J.~{Lazio}, S.~{Bale}, J.~{Bowman}, R.~{Bradley}, C.~{Carilli},
  S.~{Furlanetto}, G.~{Harker}, A.~{Loeb}, J.~{Pritchard}, {Probing the first
  stars and black holes in the early Universe with the Dark Ages Radio Explorer
  (DARE)}, Advances in Space Research 49 (2012) 433--450.
\newblock \href {http://arxiv.org/abs/1106.5194} {\path{arXiv:1106.5194}},
  \href {http://dx.doi.org/10.1016/j.asr.2011.10.014}
  {\path{doi:10.1016/j.asr.2011.10.014}}.

\bibitem{Alexander-et-al-1975-a}
J.~K. {Alexander}, M.~L. {Kaiser}, J.~C. {Novaco}, F.~R. {Grena}, R.~R.
  {Weber}, {Scientific instrumentation of the Radio-Astronomy-Explorer-2
  satellite}, Astron. Astrophys 40 (1975) 365--371.

\bibitem{NASA-Surveyor-1969-a}
NASA, {Surveyor Program Results}, no. {NASA SP-184} in NASA Special
  Publications, NASA Office of Technology Utilization, 1969.

\bibitem{Eubanks-2013-e}
T.~M. {Eubanks}, {A Space Elevator for the Far Side of the Moon}, in: Annual
  Meeting of the Lunar Exploration Analysis Group, LPI Contributions, 2013, p.
  7047.

\bibitem{Lawrence-et-al-2003-a}
D.~J. {Lawrence}, R.~C. {Elphic}, W.~C. {Feldman}, T.~H. {Prettyman},
  O.~{Gasnault}, S.~{Maurice}, {Small-area thorium features on the lunar
  surface}, Journal of Geophysical Research (Planets) 108 (2003) 6--1.
\newblock \href {http://dx.doi.org/10.1029/2003JE002050}
  {\path{doi:10.1029/2003JE002050}}.

\bibitem{Wagner-Robinson-2014-a}
R.~V. {Wagner}, M.~S. {Robinson}, {Distribution, formation mechanisms, and
  significance of lunar pits}, Icarus 237 (2014) 52--60.
\newblock \href {http://dx.doi.org/10.1016/j.icarus.2014.04.002}
  {\path{doi:10.1016/j.icarus.2014.04.002}}.

\bibitem{Fa-Jin-2007}
W.~{Fa}, Y.-Q. {Jin}, {Quantitative estimation of helium-3 spatial distribution
  in the lunar regolith layer}, Icarus 190 (2007) 15--23.
\newblock \href {http://dx.doi.org/10.1016/j.icarus.2007.03.014}
  {\path{doi:10.1016/j.icarus.2007.03.014}}.

\bibitem{Turyshev-et-al-2013-a}
S.~G. {Turyshev}, J.~G. {Williams}, W.~M. {Folkner}, G.~M. {Gutt}, R.~T.
  {Baran}, R.~C. {Hein}, R.~P. {Somawardhana}, J.~A. {Lipa}, S.~{Wang},
  {Corner-cube retro-reflector instrument for advanced lunar laser ranging},
  Experimental Astronomy 36 (2013) 105--135.
\newblock \href {http://arxiv.org/abs/1210.7857} {\path{arXiv:1210.7857}},
  \href {http://dx.doi.org/10.1007/s10686-012-9324-z}
  {\path{doi:10.1007/s10686-012-9324-z}}.

\bibitem{Williams-et-al-2006-a}
J.~G. {Williams}, S.~G. {Turyshev}, D.~H. {Boggs}, J.~T. {Ratcliff}, {Lunar
  laser ranging science: Gravitational physics and lunar interior and geodesy},
  Advances in Space Research 37 (2006) 67--71.
\newblock \href {http://arxiv.org/abs/gr-qc/0412049}
  {\path{arXiv:gr-qc/0412049}}, \href
  {http://dx.doi.org/10.1016/j.asr.2005.05.013}
  {\path{doi:10.1016/j.asr.2005.05.013}}.

\bibitem{Grun-et-al-2011-a}
E.~{Gr{\"u}n}, M.~{Horanyi}, Z.~{Sternovsky}, {The lunar dust environment},
  {Planetary and Space Science} 59 (2011) 1672--1680.
\newblock \href {http://dx.doi.org/10.1016/j.pss.2011.04.005}
  {\path{doi:10.1016/j.pss.2011.04.005}}.

\bibitem{Drolshagen-et-al-2008-a}
G.~{Drolshagen}, V.~{Dikarev}, M.~{Landgraf}, H.~{Krag}, W.~{Kuiper},
  {Comparison of Meteoroid Flux Models for Near Earth Space}, Earth Moon and
  Planets 102 (2008) 191--197.
\newblock \href {http://dx.doi.org/10.1007/s11038-007-9199-6}
  {\path{doi:10.1007/s11038-007-9199-6}}.

\bibitem{Forward-Hoyt-1995-a}
R.~Forward, R.~Hoyt, Failsafe multiline Hoytether lifetimes, American Institute
  of Aeronautics and Astronautics, 1995, pp. AlAA 95--28903.
\newblock \href {http://dx.doi.org/doi:10.2514/6.1995-2890}
  {\path{doi:doi:10.2514/6.1995-2890}}.

\bibitem{Eubanks-et-al-2015-a}
T.~M. {Eubanks}, C.~{Maccone}, C.~F. {Radley}, {Lunar Farside Radio Astronomy
  Base Facilitated by Lunar Elevator}, in: Annual Meeting of the Lunar
  Exploration Analysis Group, Vol. 1863 of LPI Contributions, 2015, p. 2014.

\bibitem{Jaumann-et-al-2013-a}
R.~{Jaumann}, H.~{Hiesinger}, M.~{Anand}, I.~A. {Crawford}, R.~{Wagner},
  F.~{Sohl}, B.~L. {Jolliff}, F.~{Scholten}, M.~{Knapmeyer}, H.~{Hoffmann},
  H.~{Hussmann}, M.~{Grott}, S.~{Hempel}, U.~{K{\"o}hler}, K.~{Krohn},
  N.~{Schmitz}, J.~{Carpenter}, M.~{Wieczorek}, T.~{Spohn}, M.~S. {Robinson},
  J.~{Oberst}, {Geology, geochemistry, and geophysics of the Moon: Status of
  current understanding}, {Planetary and Space Science} 74 (2012) 15--41.
\newblock \href {http://dx.doi.org/10.1016/j.pss.2012.08.019}
  {\path{doi:10.1016/j.pss.2012.08.019}}.

\bibitem{Jester-Falcke-2009-a}
S.~{Jester}, H.~{Falcke}, {Science with a lunar low-frequency array: From the
  dark ages of the Universe to nearby exoplanets}, New Astronomy Reviews 53
  (2009) 1--26.
\newblock \href {http://arxiv.org/abs/0902.0493} {\path{arXiv:0902.0493}},
  \href {http://dx.doi.org/10.1016/j.newar.2009.02.001}
  {\path{doi:10.1016/j.newar.2009.02.001}}.

\bibitem{Maccone-2008-a}
C.~{Maccone}, {Protected antipode circle on the Farside of the Moon}, Acta
  Astronautica 63 (2008) 110--118.

\bibitem{Suggs-et-al-2014-a}
R.~M. {Suggs}, D.~E. {Moser}, W.~J. {Cooke}, R.~J. {Suggs}, {The flux of
  kilogram-sized meteoroids from lunar impact monitoring}, Icarus 238 (2014)
  23--36.
\newblock \href {http://arxiv.org/abs/1404.6458} {\path{arXiv:1404.6458}},
  \href {http://dx.doi.org/10.1016/j.icarus.2014.04.032}
  {\path{doi:10.1016/j.icarus.2014.04.032}}.

\bibitem{Lognonne-et-al-2012-a}
P.~{Lognonne}, W.~B. {Banerdt}, K.~{Hurst}, D.~{Mimoun}, R.~{Garcia},
  M.~{Lefeuvre}, J.~{Gagnepain-Beyneix}, M.~{Wieczorek}, A.~{Mocquet},
  M.~{Panning}, E.~{Beucler}, S.~{Deraucourt}, D.~{Giardini}, L.~{Boschi},
  U.~{Christensen}, W.~{Goetz}, T.~{Pike}, C.~{Johnson}, R.~{Weber},
  K.~{Larmat}, N.~{Kobayashi}, J.~{Tromp}, {Insight and Single-Station
  Broadband Seismology: From Signal and Noise to Interior Structure
  Determination}, in: Lunar and Planetary Science Conference, Vol.~43 of Lunar
  and Planetary Inst.~Technical Report, 2012, p. 1983.

\bibitem{Ishimatsu-et-al-2015-a}
T.~Ishimatsu, O.~L. de~Weck, J.~A. Hoffman, Y.~Ohkami, R.~Shishko,
  \href{http://dx.doi.org/10.2514/1.A33235}{Generalized multicommodity network
  flow model for the earth--moon--mars logistics system}, Journal of Spacecraft
  and Rockets 53~(1) (2015) 25--38.
\newblock \href {http://dx.doi.org/10.2514/1.A33235}
  {\path{doi:10.2514/1.A33235}}.
\newline\urlprefix\url{http://dx.doi.org/10.2514/1.A33235}

\end{thebibliography}







\end{document}